\documentclass[showkeys,eqsecnum,showpacs,prd,superscriptaddress,nofootinbib]{revtex4}
\usepackage[dvips]{graphicx}
\usepackage{color}
\usepackage{amsmath}
\usepackage[dvips]{graphicx}
\begin{document}

\title{Thermodynamics of Spherically Symmetric Spacetimes in Loop Quantum Gravity}

\author{Jarmo M\"akel\"a} 

\email[Electronic address: ]{jarmo.makela@vamk.fi}  
\affiliation{Vaasa University of Applied Sciences, Wolffintie 30, 65200 Vaasa, Finland}

\begin{abstract}  

The choice of the area operator in loop quantum gravity is by no means unique. In addition to the area operator commonly used in loop quantum gravity there is also an area operator introduced by Krasnov in 1998, which gives uniformly spaced area spectra for the horizons of spacetime.  Using Krasnov's area operator we consider the thermodynamics of spherically symmetric spacetimes equipped with horizons in loop quantum gravity. Among other things, our approach implies, in a pretty simple manner, that every horizon of spacetime emits thermal radiation and possesses entropy which, in the natural units, is one-quarter of its area. When applied to the de Sitter spacetime loop quantum gravity provides an explanation both to the presence and the smallness of the cosmological constant.

\end{abstract}


\maketitle

\section{Introduction}

   For almost 30 years {\it loop quantum gravity} has been developed as a serious candidate for a quantum theory of
gravitation. {\cite{yy}} The basic ideas of loop quantum gravity are very simple, and their physical origin may be traced back 
to an almost trivial observation that the laws of physics must preserve invariant in the local rotations of the 
frame of reference of the observer. As a consequence, one may re-formulate Einstein's general theory of relativity
as if it were a gauge field theory with the rotation group $SO(3)$ as one of its 
symmetry groups. Applying the standard rules of quantization to the resulting gauge field theory one obtains a 
quantum theory of gravitation. {\cite{kaa}}

    The quantum states of the gravitational field in loop quantum gravity are the so-called 
{\it spin network states}. {\cite{koo}} In very broad terms, a spin network in loop quantum gravity may be described as a 
graph lying on a spacelike hypersurface of spacetime such that every edge of the graph is associated with a 
$(2j+1)$-dimensional  irreducible representation $(j = 0, 1/2, 1, 3/2,...)$ of the group $SU(2)$, whose generators obey exactly the same Lie algebra as the generators of the group $SO(3)$. The most important single 
prediction of loop quantum gravity is that the area of any spacelike two-surface has a discrete spectrum. More
precisely, it turns out that the possible area eigenvalues of a given spacelike two-surface are, in the SI
units, of the form: {\cite{koo}}
\begin{equation}
A = 8\pi\gamma\ell_{Pl}^2\sum_p\sqrt{j_p(j_p+1)},
\end{equation}
where $\ell_{Pl} := \sqrt{\hbar G/c^3}\approx 1.6\times 10^{-35}m$ is the Planck length, and $\gamma$, which is
a pure number of the order of unity, is the so-called Immirzi parameter. In Eq. (1.1) we have summed over the 
so-called {\it punctures} $p$ of the spin network on the two-surfaces, or the edges, which intersect the 
two-surface. The quantum number $j_p$ determines the dimension of the representation of the group $SU(2)$ 
associated with the puncture $p$.

  Eq. (1.1) is a standard result of loop quantum gravity. However, it was pointed out already in the year 1998
 by Krasnov {\cite{nee}} that it is possible to define the area operator in loop quantum gravity in such a way that the 
area eigenvalues of a given spacelike two-surface of spacetime are not given by the right hand side of Eq. (1.1),
but by a somewhat more simple formula:
\begin{equation}
A = 8\pi\gamma\ell_{Pl}^2\sum_p j_p.
\end{equation}
A disadvantage of the area operator producing the area spectrum in Eq. (1.2) is that it does not possess the 
local $SO(3)$ invariance. Nevertheless, it was argued by Krasnov that if spacetime has boundaries --horizons, for
example-- the $SO(3)$ symmetry does  not necessarily have to be conserved on the boundaries. As a consequence, the area spectrum of a 
spacelike two-surface very close to a horizon of spacetime might well be given by Eq. (1.2), rather than by 
Eq. (1.1). An advantage of Eq. (1.2) is that it predicts uniformly spaced area spectra for the horizons of 
spacetime. Indeed, there is compelling evidence to the effect that the event horizons of black holes, for
example, have a uniform spacing in their area spectra. {\cite{vii}}

    In this paper we shall take Eq. (1.2) as a starting point. More precisely, we shall assume that whenever 
we pick up a spacelike two-surface sufficiently close to a horizon of spacetime, its area spectrum is given by 
Eq. (1.2). Using this assumption we shall consider the statistical physics and the thermodynamics of spacetimes
equipped  with horizons. To make things simple we shall restrict our attention to spherically symmetric 
spacetimes, or spacetimes, where the line element may be written as:
\begin{equation}
ds^2 = -f(r,\alpha_1,...,\alpha_s)\,dt^2 + \frac{dr^2}{f(r,\alpha_1,...,\alpha_s)} + r^2\,d\theta^2 
+ r^2 \sin^2(\theta)\,d\phi^2.
\end{equation}
In this equation $f(r,\alpha_1,...,\alpha_s)$ is assumed to be a smooth function of the radial coordinate $r$, 
and of $s$ parameters $\alpha_1, \alpha_2,..., \alpha_s$. The parameters $\alpha_1,...,\alpha_s$ may be, for 
instance, the mass, the cosmological constant, electric charge, etc. We shall see that employing Eq. (1.2) it 
is possible to obtain, as in Refs. {\cite{kuu, seite, kasi}}, an explicit, analytic expression for the partition function of 
spacetime. Using this partition function and choosing
\begin{equation}
\gamma = \frac{\ln(2)}{\pi}
\end{equation}
in Eq. (1.2) one finds, among other things, that every horizon of spacetime emits thermal radiation and possesses 
entropy which, in the natural units, is one-quarter of its area.

   Of particular interest are the predictions obtainable from Eq. (1.2) for the properties of the cosmological 
constant. When applying the general results to be derived in Sections 2-8 to the de Sitter spacetime one 
discovers, treating the cosmological "constant" not really as a constant, but as a free thermodynamical variable,
that in very low temperatures the cosmological constant takes a huge, Planck-size value which, in the SI units,
is around $10^{87}s^{-2}$. However, at a certain temperature the de Sitter spacetime performs a {\it phase
transition}, where the cosmological constant drops drastically. After the phase transition has been completed,
the cosmological constant takes the form:
\begin{equation}
\Lambda = \frac{3\pi}{2\ln(2)}\frac{1}{N}\frac{c^2}{\ell_{Pl}^2},
\end{equation}
where $N$ is the number of the punctures of the spin network on a closed, spacelike two-sphere just inside of the 
cosmological horizon of the de Sitter spacetime. Since $N$ is finite, the cosmological constant is bounded from 
the below. However, since $N$ is presumably very large, the cosmological constant must be very small. Taking
$N \sim 10^{122}$ one finds that $\Lambda \sim 10^{-35} s^{-2}$, which is the present estimate, based on the
observations, for the cosmological constant. Hence it appears that loop quantum gravity may provide an explanation
both to the presence and the smallness of the cosmological constant.

   Unless otherwise stated, we shall always use the natural units, where 
$\hbar = c = G = k_B = 4\pi\epsilon_0 = 1$. 

\section{Area Operators in Loop Quantum Gravity}

One of the basic objects of loop quantum gravity is the so-called {\it densitized triad}
\begin{equation}
\tilde{E}_k^a  := \sqrt{q}\,E_k^a.
\end{equation}
In Eq. (1.2) $q$ is the determinant of the metric tensor $q_ {ab}$ $(a, b = 1, 2, 3)$ induced on the spacelike hypersurface $\Sigma$ of spacetime, where the time coordinate $t = constant$. The quantities $E_k^a$ $(a, k = 1, 2, 3)$ are the components of the triads on $\Sigma$. In other words, they have a  property
\begin{equation}
E^a_k E^b_l q_{ab} = \delta_{kl}
\end{equation}
for all $k, l = 1, 2, 3$. By means of the densitized triads one may write the area of an arbitrary two-dimensional hypersurface $S^{(2)}$ of $\Sigma$ as:
\begin{equation}
A = \int_{S^{(2)}}\sqrt{n_a n_b\tilde{E}^{ka}\tilde{E}^b_k}\,d\cal{A}.
\end{equation}
In Eq. (2.3) $n^a$ is the unit normal vector field of $S^{(2)}$ on $\Sigma$. $d\cal{A}$ is the integration measure in the parameter space determining $S^{(2)}$. In the special case, where the coordinates $x^a$ on $\Sigma$ have been chosen such that 
$x^3 = constant$ on $S^{(2)}$, the only non-zero component of $n_a$ is $n_3 = 1$, and we may write Eq. (2.3) as:
\begin{equation}
 A = \int_{S^{(2)}}\sqrt{\tilde{E}^{k3}\tilde{E}_k^3}\,dx^1\,dx^2.
\end{equation}

   The real symmetry group of general relativity is the diffeomorphism group, or the group of diffeomorphic coordinate transformations in spacetime. However, if a canonical formulation of general relativity is constructed using the densitized triads, instead of the components of the metric tensor $q_{ab}$ induced on $\Sigma$, as the fundamental objects, yet another symmetry group is brought into the theory. That group is the {\it rotation group SO(3)}. The $SO(3)$ symmetry of the resulting theory stems from the fact that when the metric tensor is written in terms of the triads $E^k_a$ as:
\begin{equation}
q_{ab} = E^k_a E_{kb},
\end{equation}
then $q_{ab}$  remains unchanged in the rotations of the triads.

      When quantizing a theory formulated in terms of the densitized triads $\tilde{E}^a_k$ one turns the densitized triads $\tilde{E}^a_k$  to the corresponding operators $\hat{\tilde{E}}^a_k$ operating on a specific state space, which is spanned by the so-called {\it spin network states}. The precise definition of spin networks used in quantum gravity may be found, for example, in Ref. {\cite{koo}}. At this point it is sufficient to say that, in broad terms, spin network may be described as a graph lying on $\Sigma$ such that every edge of the graph is associated with some irreducible representation of the group $SU(2)$, which has the same Lie algebra as the group $SO(3)$. The result, which is of particular interest to us in here states that whenever an edge of a spin network passes through $S^{(2)}$ at some point $p$, which is known as a {\it puncture} of the spin network on $S^{(2)}$, the operator $n_a\hat{\tilde{E}}_k^a$ at the puncture $p$ acts on the spin network state in the same way as does the operator 
\begin{equation}
\hat{E}_k(p) := 8\pi\gamma\hat{J}_k(p)
\end{equation}
on an element of the representation space, associated with the edge passing through $S^{(2)}$ at $p$, of the group $SU(2)$. In Eq. (2.6) $\hat{J}_k(p)$ is an  operator representing the (complexified) generator $J_k$ $(k = 1,2,3)$ of $SU(2)$ at $p$. As a consequence, the operators $\hat{J}_k(p)$ obey the commutation relations:
\begin{equation}
[\hat{J}_k(p),\hat{J}_l(p)] = i\epsilon_{kls}\hat{J}_s(p).
\end{equation}
The number $\gamma$ is known as the {\it Immirzi parameter}, and it is a pure number of the order of  unity.

   Eqs. (2.3) and (2.6) imply that the area operator takes, in loop quantum gravity, the form:
\begin{equation}
\hat{A} = 8\pi\gamma\sum_p\sqrt{\hat{J}^k(p)\hat{J}_k(p)}.
\end{equation}
We have replaced the integral by a sum, because the operators $n_a\hat{\tilde{E}}_k^a$ act on the  edges puncturing the two-surface $S^{(2)}$ only, and we have summed over the punctures $p$. According to Eq. (2.8) the area eigenvalues of a spacelike two-surface in loop quantum gravity are of the form:
\begin{equation}
A = 8\pi\gamma\sum_p\sqrt{j_p(j_p+1)}.
\end{equation}
In Eq. (2.9) $j_p$ is an integer or half-integer, and it determines the dimensionality of the representation space of the group $SU(2)$ at the edge puncturing $S^{(2)}$ at $p$ such that the dimension of the representation is $2j_p +1$.

   Eq. (2.9), which gives the area spectrum, is the most important single result produced by loop quantum gravity. No doubt the area operator $\hat{A}$ in Eq. (2.8) has many desirable properties: Its spectrum is non-negative, and it remains invariant in the local rotations of the triads at the punctures. Nevertheless, it is certainly not the only possible choice for the area operator in loop quantum gravity. For instance, it was an idea of Krasnov to consider a quantity
\begin{equation}
A_S := \int_{S^{(2)}}\tilde{E}_k^3r^k\,dx^1\,dx^2
\end{equation}
on $S^{(2)}$. {\cite{nee}} (We have chosen the coordinates on $\Sigma$ such that $x^3=constant$ on $S^{(2)}$.) In Eq. (2.10) $r^k$ is an arbitrary unit vector field on $S^{(2)}$. One observes that if the triads are rotated such that $E_3^a$ is parallel to $r^k$ everywhere on $S^{(2)}$, then $A_S$ takes its maximum, which equals with the area of $S^{(2)}$. When the triads are chosen arbitrarily, and the densitized triads $\tilde{E}_k^a$ are replaced by the corresponding operators $\hat{\tilde{E}}_k^a$ we find, using Eq. (2.6), that the quantity $A_S$ turns to an operator
\begin{equation}
\hat{A}_S := 8\pi\gamma\sum_p\hat{J}_k(p)r^k(p),
\end{equation}
which has the eigenvalues:
\begin{equation}
A_S = 8\pi\gamma\sum_p m_p,
\end{equation}
where, for arbitrary puncture $p$, $m_p \in\lbrace -j_p, -j_p +1,..., j_p -1, j_p\rbrace$.

     It must be admitted that the operator $\hat{A}_S$ defined in Eq. (2.11)  is very problematic: In contrast to the operator $\hat{A}$ in Eq. (2.8), the operator $\hat{A}_S$ is not $SO(3)$ invariant, and it may take negative, as well as positive eigenvalues. According to Krasnov, however, both of these problems may be avoided: If the region of space under consideration possesses a {\it boundary} -- a horizon, for instance -- then the $SO(3)$ symmetry is not necessarily conserved on the boundary, and we may use the operator $\hat{A}_S$ as a legitimate area operator on that boundary. Moreover, the area of that boundary is not actually given by the quantity $A_S$ in Eq. (2.10) itself, but rather by its {\it maximum}. As an analog of this classical result, the correct area eigenvalues of the boundary are given by the maximal eigenvalues of $\hat{A}_S$. (For more details of this argument, see Ref. {\cite{nee}}) Since the maximum of $m_p$ for given $j_p = 0, 1/2, 1, 3/2,...$ is $j_p$, the area eigenvalues are of the form:
\begin{equation}
A_S = 8\pi\gamma\sum_p j_p.
\end{equation}
As one may observe from Eq. (2.9), the spectrum of $\hat{A}_S$ agrees with the spectrum of $\hat{A}$ in the limit of large quantum numbers $j_p$. 

   Because the quantum numbers $j_p$ are all either integers or half-integers, Eq. (2.13) implies, when applied to the horizons of spacetime, that the horizons have uniformly spaced area spectra. There is compelling evidence for the uniform spacing of the black hole event horizon area spectrum, {\cite{vii}} and in this regard Eq. (2.13) provides a desirable result. Assuming that there are $N$ punctures on the given horizon of spacetime, we may write Eq. (2.13), in the SI units, as:
\begin{equation}
A_S = 4\pi\gamma\ell_{Pl}^2(n_1 + n_2 +...+n_N),
\end{equation}
where
\begin{equation}
\ell_{PL} := \sqrt{\frac{\hbar G}{c^3}} \approx 1.6\times 10^{-35}m
\end{equation}
is the Planck length.  We have defined the non-negative integers $n_p$ such that:
\begin{equation}
n_p := 2j_p
\end{equation}
for all $p = 1,2,..., N$.

    In this paper we shall henceforth always assume that the area eigenvalues of the horizons of spacetime obey Eq. (2.14), and thus agree with those of Krasnov's area operator. The horizon area spectrum given by Eq. (2.14) has implications, which have been investigated by the author in several papers. {\cite{kuu,seite,kasi}} However, in this paper those implications will be studied in a general and systematic fashion, emphasizing the origin of Eq. (2.14) in loop quantum gravity. For the Immirzi parameter $\gamma$ in Eq. (2.14) we shall pick up a value
\begin{equation}
\gamma = \frac{\ln(2)}{\pi},
\end{equation}
and it turns out that with this choice one may recover, beginning from Eq. (2.14), the well-known results concerning, among other things, the entropy and the radiation of black holes.

\section{Stretched and Shrinked Horizons}

 In this paper we shall consider spacetimes, where the line element may be written as:
\begin{equation}
ds^2 = -f(r,\alpha_1,...,\alpha_s)\,dt^2 + \frac{dr^2}{f(r,\alpha_1,...,\alpha_s)} + r^2\,d\theta^2 + r^2\sin^2(\theta)\,d\phi^2.
\end{equation}
In this equation $r$, $\theta$ and $\phi$ are the spherical coordinates, and the function $f(r,\alpha_1,...,\alpha_s)$ is assumed to be a smooth function of the radial coordinate $r>0$, and of the parameters $\alpha_j$ $(j = 1, 2,..., s)$. The parameters $\alpha_j$ may be, for instance, the mass, the electric charge, and the cosmological constant. We shall also assume that for some values $r_k$ $(k = 1, 2,..., s)$ of the radial coordinate $r$, which depend on the parameters $\alpha_j$, the spacetime has {\it horizons}, where
\begin{equation}
f(r_k,\alpha_1,...,\alpha_s) = 0.
\end{equation}
At the horizons the radial derivative of $f$ is assumed to be non-vanishing, i.e. 
\begin{equation}
\frac{\partial f}{\partial r}\vert_{r=r_k} \ne 0
\end{equation}
for all $k = 1, 2,..., s$.

    We shall begin our investigation of the properties of spacetime equipped with the line element (3.1) by means of a study of an observer at rest with respect to the coordinates $r$, $\theta$ and $\phi$. The only non-vanishing component of the future pointing unit tangent vector field of the congruence of the world lines of such observers is:
\begin{equation}
u^t = f^{-1/2}.
\end{equation}
As a consequence, the only non-vanishing component of the proper acceleration vector field 
\begin{equation}
a^\mu := u^\nu u^\mu_{;\nu} = u^\nu(u^\mu_{,\nu} + \Gamma_{\nu\alpha}^\mu u^\alpha)
\end{equation}
of the congruence is
\begin{equation}
a^r = u^t\Gamma_{tt}^r u^t = \frac{1}{2}\frac{\partial f}{\partial r}.
\end{equation}
The norm of this vector field is:
\begin{equation}
a := \sqrt{a^r a_r} = \frac{1}{2}f^{-1/2}\vert\frac{\partial f}{\partial r}\vert,
\end{equation}
and it gives the proper acceleration of an observer at rest with respect to the coordinates $r$, $\theta$ and $\phi$.

    Since $f = 0$ and $\frac{\partial f}{\partial r}\ne 0$ at the horizons, Eq. (3.7) implies that the proper acceleration $a$ is singular at the horizons. Assuming that spacetime has a horizon, where $r  = r_k$ we find that if 
\begin{equation}
\frac{\partial f}{\partial r}\vert_{r=r_k} > 0,
\end{equation}
then $f$ will change its sign from negative to positive at the horizon, whereas if 
\begin{equation}
\frac{\partial f}{\partial r}\vert_{r=r_k} < 0,
\end{equation}
$f$ will change its sign at the horizon from positive to negative. To avoid the singularity in $a$  at the horizon, when Eq. (3.8) holds we shall investigate, instead of the horizon itself, the so-called {\it stretched horizon}, or a spacelike two-sphere, where $r = constant > r_k$ just outside of the horizon, where $r = r_k$. Consequently, if Eq. (3.9) holds, we shall consider the so-called {\it shrinked horizon}, which is a spacelike two-sphere just {\it inside} of the horizon.

    The parameters $\alpha_j$ may change in the thermodynamical processes. For instance, if the line element (3.1) describes a spherically symmetric spacetime with a Schwarzschild black hole, the only parameter is the Schwarzschild mass $M$ of the hole, and when the black hole evaporates, $M$ will decrease. As a consequence, the location of the event horizon of the Schwarzschild black hole will also change. No matter, whether we consider stretched or shrinked horizon of spacetime, we shall always assume that when the parameters $\alpha_j$ are changed, the radius of the stretched or shrinked horizon will also change, but in such a way that the proper acceleration $a$, as such as given in Eq. (3.7), remains as a constant. We shall consider the thermodynamical processes of spherically symmetric spacetimes from the point of view of an observer lying either on a stretched or a shrinked horizon of spacetime, where $a = constant$. The reason for this decision of ours is that observers with the same proper acceleration may be considered in some sense equivalent, and hence the processes detected by such observers reflect the changes in the real physical properties of spacetime, rather than the properties of the frame of reference used by the observer.

    When $a = constant$, we must have:
\begin{equation}
da = \frac{\partial a}{\partial r}\,dr + \frac{\partial a}{\partial \alpha_1}\,d\alpha_1 + ... +\frac{\partial a}{\partial \alpha_s}\,d\alpha_s = 0
\end{equation}
for the infinitesimal changes $dr$ and $d\alpha_k$ in the radius $r$ of the stretched or shrinked horizon, and in the parameters $\alpha_k$. Using Eq. (3.7) we find:
\begin{equation}
\frac{\partial f}{\partial r}\lbrack \frac{\partial f}{\partial r}\,dr + \frac{\partial f}{\partial \alpha_1}\,d\alpha_1 +...+ \frac{\partial f}{\partial \alpha_s}\,d\alpha_s\rbrack = 2f\lbrack \frac{\partial^2 f}{\partial r^2}\,dr +...+ \frac{\partial^2 f}{\partial \alpha_1\partial r}\,d\alpha_1 +...+ \frac{\partial^2 f}{\partial \alpha_s \partial r}\,d\alpha_s\rbrack,
\end{equation}
no matter, whether the horizon is stretched or shrinked. Since $f \longrightarrow 0$, when the stretched or shrinked horizon gets close to the horizon itself, whereas $\frac{\partial f}{\partial r} \ne 0$, we must have, close to the horizon:
\begin{equation}
\frac{\partial f}{\partial r}\,dr + \frac{\partial f}{\partial \alpha_1}\,d\alpha_1 +...+ \frac{\partial f}{\partial \alpha_s}\,d\alpha_s = 0.
\end{equation}
Eq. (3.12) may be written as:
\begin{equation}
df = 0,
\end{equation}
where $df$ is the total differential of $f$. Since $f = 0$ at the horizon, Eq. (3.13) implies that a stretched or shrinked horizion originally close to the horizon will stay close to the horizon, no matter what may happen to the parameters $\alpha_k$. In this sense our concepts of stretched and shrinked horizons are well-defined.

 \section{Energy}

The concept of energy plays a key role in the thermodynamical investigation of all systems. The concept of energy frequently used in general relativity is the so-called {\it Brown-York energy}: {\cite{ysi}}
\begin{equation}
E_{BY} := -\frac{1}{8\pi}\oint_{S^{(2)}}(k - k_0)\,dA.
\end{equation}
In Eq. (4.1) $k$ is the trace of the exterior curvature tensor induced on a closed, spacelike two-surface $S^{(2)}$ embedded into a spacelike hypersurface $\Sigma$ of spacetime, where the time coordinate $t = constant$. $k_0$ is the trace of the exterior curvature tensor, when the two-surface $S^{(2)}$ has been embedded in flat space. $dA$ is the area element on the two-surface, and we have integrated over the whole two-surface.

    For stationary spacetimes the Brown-York energy may be understood as the energy of the gravitational field inside of the closed two-surface $S^{(2)}$. When the line element of spacetime is written as in Eq. (3.1), the only non-zero components of the exterior curvature tensor on the closed two-sphere, where $r = constant$ are, whenever $f > 0$:
\begin{subequations}
\begin{eqnarray}
k_{\theta\theta} &=& - f^{-1/2}\Gamma_{\theta\theta}^r = rf^{1/2},\\
k_{\phi\phi} &=& -f^{-1/2}\Gamma_{\phi\phi}^r = r\sin^2(\theta)f^{1/2},
\end{eqnarray}
\end{subequations}
and its trace is:
\begin{equation}
k = k_\theta^{\,\,\theta} + k_\phi^{\,\,\phi} = \frac{2}{r}f^{1/2}.
\end{equation}
In the flat space $f=1$, and the trace of the exterior curvature tensor is:
\begin{equation}
k_0 = \frac{2}{r}.
\end{equation}
The area of the two-sphere, where $r = constant$, is:
\begin{equation}
A = 4\pi r^2,
\end{equation}
and hence the Brown-York energy takes the form:
\begin{equation}
E_{BY} = r(1 - f^{1/2}).
\end{equation}

  Consider now, what happens to the Brown-York energy $E_{BY}$, when the parameters $\alpha_k$ take on infinitesimal changes. The resulting change in $E_{BY}$ is:
\begin{equation}
dE_{BY} = -\frac{1}{2}rf^{-1/2}(\frac{\partial f}{\partial \alpha_1}\,d\alpha_1 + ... + \frac{\partial f}{\partial \alpha_s}\,d\alpha_s).
\end{equation}
When we are either on a stretched or a shrinked horizon, where the proper acceleration $a = constant$, the infinitesimal changes $d\alpha_k$ correspond to the change $dr$ in the radius $r$ of the horizon such that Eq. (3.12) holds. Using Eqs. (4.7) and (3.12) we find:
\begin{equation}
dE_{BY} = \frac{1}{2}f^{-1/2}\frac{\partial f}{\partial r}r\,dr.
\end{equation}
For stretched horizon $\frac{\partial f}{\partial r} > 0$, and Eq. (3.7) implies:
\begin{equation}
dE_{BY} = ar\,dr.
\end{equation}
Eq. (4.9) gives the change in the Brown-York energy from the point of view of an observer on a stretched horizon of spacetime, if the parameters $\alpha_k$, and the radius $r$ of the stretched horizon are changed in such a way that the proper acceleration $a$ is kept as a constant during the process. The change $dr$ results in the change 
\begin{equation}
dA = 8\pi r\,dr
\end{equation}
in its area, and so we may write Eq. (4.9) as:
\begin{equation}
dE_{BY} = \frac{a}{8\pi}\,dA.
\end{equation}
Since the proper acceleration $a$ is kept as a constant, we may thus conclude that in a process, where the area of the stretched horizon is increaseed from zero to $A$, the energy carried inside of the stretched horizon has increased from zero to 
\begin{equation}
E = \frac{a}{8\pi}A.
\end{equation}
For all practical purposes we may identify the area $A$ of the stretched horizon with the area of the horizon itself. The energy given by Eq. (4.12) may be understood as the amount of energy carried, from the point of view of an observer with constant proper acceleration $a$, inside of the horizon during the creation of the horizon. For instance, if a Schwarzschild black hole is created by means of a gravitational collapse, mass flows past of the observer, and Eq. (4.12) gives the energy associated with that mass from the point of view of the observer.

    For shrinked horizons the situation is a bit more complicated. Since $\frac{\partial f}{\partial r} < 0$ for shinked horizons, Eq. (3.7) implies that Eq. (4.11) must be replaced by an equation:
\begin{equation}
dE_{BY} = -\frac{a}{8\pi}\,dA.
\end{equation}
The presence of the minus sign on the right hand side of Eq. (4.13) may be understood on grounds of the fact that since $\frac{\partial f}{\partial r} < 0$, $f$ will change its sign from positive to negative at the horizon. As a consequence, energy flows from inside to the {\it outside} of the horizon during the process, where the horizon is created, and the minus sign on the right hand side of Eq. (4.13) indicates that the amount of energy inside of the horizon has {\it decreased} during its creation. Indeed, when $\frac{\partial f}{\partial r} < 0$, Eq. (3.6) implies that the radial component $a^r$ of the proper acceleration vector of an observer at rest with respect to the coordinates $r$, $\theta$ and $\phi$ is negative, which means that material particles in a free fall are carried from the inside to the outside  of the horizon. The energy carried outside of the horizon during its creation is
\begin{equation}
E = \frac{a}{8\pi}A,
\end{equation}
and it may be identified as the gravitational energy perceived by the observer. In general, the gravitational energy perceived by an observer with proper acceleration $a_k$ is 
\begin{equation}
E_k = \frac{a_k}{8\pi}A_k
\end{equation}
for any horizon with area $A_k$, no matter, whether the horizon is stretched or shrinked. 

\section{The Model}

In this paper we shall assume that in every stretched or shrinked horizon of spherically symmetric spacetime the spin network has a fixed, finite number of punctures. Denoting the number of punctures on a stretched or shrinked horizon $k$ $(k = 1, 2,..., s)$ by $N_k$ Eqs. (2.14) and (2.17) imply that the area of the horizon is of the form:
\begin{equation}
A_k = 4\ln(2)\ell_{Pl}^2(n_{k1} + n_{k2} +...+ n_{kN_k}),
\end{equation}
where the non-negative integer $n_{kp}$ $(p = 1, 2, 3,..., N_k)$ determines, in the units of $4\ln(2)\ell_{Pl}^2$, the area contributed by the puncture $p$ to the horizon $k$. We shall assume that the spin network has on the horizon 1 at most as many punctures as it does on on the horizon 2 which, in turn, has at most as many puncture as does the horizon 3, and so on. In other words, we shall assume that
\begin{equation}
N_1 \le  N_2 \le N_3 \le...\le N_s.
\end{equation}
We shall also assume that the areas $A_k$ of the shrinked horizons have the property:
\begin{equation}
A_1 \le A_2 \le...\le A_s.
\end{equation}
This means that the horizons cannot get past of each other. This condition is satisfied, if we assume that between the quantum numbers $n_{kp}$ and $n_{(k+1)p}$ there is the relationship:
\begin{equation}
n_{(k+1)p} = n_{kp} + m_{(k+1)p}
\end{equation}
for every $k = 1, 2,..., (s-1)$ and $p = 1, 2,..., N_k$. 
In Eq. (5.4) the quantities $m_{(k+1)p}$ are non-negative integers.

     Eqs. (5.1), (5.2) and (5.4) imply:
\begin{subequations}
\begin{eqnarray}
A_1 &=& 4\ln(2)\ell_{Pl}^2(n_{11} + n_{12} +...+ n_{1N_1}),\\
A_2 &=& 4\ln(2)\ell_{Pl}^2[(n_{11} + m_{21}) +...+ (n_{1N_1} + m_{2N_1}) + n_{2(N_1+1)} +...+ n_{2N_2})],\\
A_3 &=& 4\ln(2)\ell_{Pl}^2[(n_{21} + m_{31}) +...+ (n_{2N_2} + m_{3N_2}) + n_{3(N_2 +1)} +...+ n_{3N_3}],
\end{eqnarray}
\end{subequations}
and so on. Another way of writing Eq. (5.5c) is:
\begin{eqnarray}
A_3 = 4\ln(2)\ell_{Pl}^2&[&(n_{11} + m_{21} + m_{31}) + (n_{12} + m_{22} + m_{32}) +...+ (n_{1N_1} + m_{2N_1}
+ m_{3N_1})\cr
         &+&(n_{2(N_1 +1)} + m_{3(N_1 +1)}) + (n_{2(N_1+2)} + m_{3(N_1+2)}) +...+ (n_{2N_2} + m_{3N_2})\cr
         &+&n_{3(N_2+1)} + n_{3(N_2+2)} +...+ n_{3N_3}].
\end{eqnarray}
Observing Eqs. (5.5) and (5.6) we find that $A_1$ is determined by $N_1$ quantum numbers, $A_2$ by $N_1 + N_2$ quantum numbers, and $A_3$ by $N_1 + N_2 + N_3$ quantum numbers. In general, the area $A_k$ of horizon $k$ is determined by $N_1 + N_2 +...+ N_k$ quantum numbers for all $k = 1, 2, 3,..., s$. According to Eq. (4.15) the energy of the gravitational field perceived by an observer on the stretched or shrinked horizon $k$ depends on the area $A_k$ of that horizon only. Hence it follows that the number of independent quantum numbers determining the energy from the point of view of an observer on the horizon 1 equals with the number of the punctures of the spin network on that horizon. On the horizon 2, however, the number of quantum numbers determining the energy equals with the total number of punctures on {\it both} of the horizons 1 and 2. Finally, the number of quantum numbers determining the energy from the point of view of an observer on the outermost horizon $s$ equals with the total number of the punctures of the spin network on {\it all} horizons of spacetime. So we find that in our model the quantum-mechanical degrees of freedom do not lie, even from the point of view of an observer on the outermost horizon, on the outermost horizon only but, in some sense, on all horizons of spacetime.  

\section{Partition Function}

    In general, the partition function of any system is:
\begin{equation}
Z(\beta) = \sum_n g(E_n)e^{-\beta E_n}.
\end{equation}
In Eq. (6.1) we have summed over the possible energies $ E_n$ of the system. $g(E_n)$ is the number of the degenerate states associated with the same energy $E_n$ of the system, whereas $\beta := 1/T$ is the temperature parameter of the system.

In our model the possible energies perceived by an observer on the stretched or shrinked horizon of spherically symmetric spacetime are, according to Eqs. (4.15), (5.1) and (5.4), of the form:
\begin{equation}
E_{n_k} = \frac{a_k}{2\pi}\ln(2)n_k,
\end{equation}
where the positive integer $n_k$ is a sum of $N_1 + N_2 +...+ N_k$ non-negative integers. In our model we take the number $g(E_{n_k})$ of the degenerate states associated with the same energy to equal with the number of ways of expressing $n_k$ as a sum of at most $N_1 + N_2 +...+N_k$ non-negative integers. More precisely, we take $g(E_{n_k})$ to equal with the number of ordered strings $(q_1,q_2,...,q_l)$ of positive integers $q_1, q_2,..., q_l$ $(l\le N_1 + N_2 +...+ N_k)$ such that
\begin{equation}
q_1 + q_2 +...+ q_l = n_k,
\end{equation}
and $l \le n_k$. The number of ways of expressing a given positive integer $n_k$ as a sum of $l$ positive integers equals with the number of ways of arranging $n_k$ identical balls in a row such that there are $l - 1$ divisions in the $n_k - 1$ empty spaces between the balls. There are
\begin{equation}
\left(\begin{array}{cc} n_k-1\\
l-1\end{array}\right)
= \frac{(n_k - 1)!}{(l-1)!(n_k - l)!}
\end{equation}
ways of choosing the positions of the divisions, and therefore Eq. (6.4) gives the possible number of ways of expressing $n_k$ as a sum of $l$ positive integers Hence it follows that the number of degenerate states associated with the same energy $E_{n_k}$ is
\begin{equation}
g(E_{n_k}) = \sum_{l=1}^{{\cal N}_k}\left(\begin{array}{cc}n_k - 1\\l-1\end{array}\right),
\end{equation}
whenever
\begin{equation}
n_k \ge {\cal N}_k := N_1 + N_2 +...+ N_k.
\end{equation}
In the special case, where  $n_k = {\cal N}_k$, we have:
\begin{equation}
g(E_{n_k}) = \sum_{l=1}^{n_k}\left(\begin{array}{cc}n_k - 1\\l - 1\end{array}\right) = 2^{n_k - 1}.
\end{equation}
When $n_k < {\cal N}_k$, $g(E_{n_k})$ is simply the number of ways of writing the positive integer $n_k$ as a sum of positive integers, no matter how many. That number is given by the right hand side of Eq. (6.7), and so we find that $g(E_{n_k})$ is always given by Eq. (6.7), when $n_k < {\cal N}_k$. 

    Eqs. (6.1), (6.2), (6.5) and (6.7) imply that the partition function
\begin{equation}
Z(\beta_k) = \sum_{n_k}g(E_{n_k})e^{-\beta_k E_{n_k}}
\end{equation}
of spacetime from the point of view of an observer on a stretched or shrinked horizon $k$ of spacetime, where the proper acceleration $a_k = constant$ is of the form:
\begin{equation}
Z(\beta_k) = Z_1(\beta_k) + Z_2(\beta_k),
\end{equation}
where 
\begin{subequations}
\begin{eqnarray}
Z_1(\beta_k) &=& \frac{1}{2}\sum_{n_k = 1}^{{\cal{N}}_k} 2^{(1-\beta_k T_{Ck})n_k},\\
Z_2(\beta_k) &=&\sum_{n_k={\cal N}_k}^\infty \left[\sum_{j=0}^{{\cal N}_k}\left(\begin{array}{cc}n_k - 1\\j\end{array}\right)2^{-n_k\beta_k T_{Ck}}\right].
\end{eqnarray}
\end{subequations}
In these equations we have defined the temperature
\begin{equation}
T_{Ck} := \frac{a_k}{2\pi},
\end{equation}
which agrees with the {\it Unruh temperature} {\cite{kymmenen}} measured for a thermal radiation by an observer with proper acceleration $a_k$. The temperature $T_{Ck}$ plays an important role in our discussion. We shall therefore call $T_{Ck}$ as the {\it characteristic temperature} of the horizon $k$.

   The expression in Eqs. (6.9) and (6.19) for the partition function of spacetime from the point of view of an observer on a horizon $k$ of spacetime is almost identical to the partition function obtained in Ref. {\cite{kuu}} for the Schwarzschild black hole. The only difference of Eqs. (6.10a) and (6.10b), when compared with the corresponding equations (55) and (56) in Ref. {\cite{kuu}} is that in Eqs. (6.10a) and (6.10b) the positive integer $N$ in Ref. {\cite{kuu}} has been replaced by ${\cal N}_k$. As a consequence, the final expression for the partition function may be obtained, quite simply, by means of a replacement of $N$ by ${\cal N}_k$ in Eqs. (61) and (62) in Ref. {\cite{kuu}}. So we find that the partition function $Z(\beta_k)$ in Eq. (6.7) may be calculated explicitly, and it takes the form:
 \begin{equation}
Z(\beta_k) = \frac{1}{2^{\beta_k T_ {Ck}} - 2}\left[1 - \left(\frac{1}{2^{\beta_k T_{Ck}} - 1}\right)^{{\cal N}_k + 1}\right],
\end{equation}
when $\beta_k T_{Ck} \ne 1$ and the form:
\begin{equation}
Z(\beta_k) ={\cal N}_k + 1,
\end{equation}
when $\beta_k T_{Ck} = 1$. 

\section{Phase Transition}

Eq. (6.12) implies that the average energy 
\begin{equation}
E_k(\beta_k) = -\frac{\partial}{\partial\beta_k}\ln Z(\beta_k)
\end{equation}
of spacetime and the matter fields from the point of view of an observer residing on a stretched or shrinked horizon $k$ of spacetime takes the form:
\begin{equation}
E_k(\beta_k) = E_k^{(1)}(\beta_k) + E_k^{(2)}(\beta_k),
\end{equation}
where we have denoted:
\begin{subequations}
\begin{eqnarray}
E_k^{(1)}(\beta_k) :&=& \frac{2^{\beta_k T_{Ck}}}{2^{\beta_k T_{CK}} - 2}T_{Ck}\ln(2),\\
E_k^{(2)}(\beta_k) :&=& \frac{2^{\beta_k T_{Ck}}}{2^{\beta_k T_{Ck}} - 1 - (2^{\beta_k T_{Ck}} - 1)^{{\cal N}_k +2}}({\cal N}_k + 1)T_{Ck}\ln(2).
\end{eqnarray}
\end{subequations}
We may assume that the number ${\cal N}_k$, which gives the number of the punctures of the spin network on all horizon up to the horizon $k$, to be very large, indeed. One observes that the quantity
\[
(2^{\beta_k T_{Ck}} - 1)^{{\cal N}_k + 2}
\]
depends critically, in the large ${\cal N}_k$ limit, on whether $\beta_k T_{Ck}$ is greater or less than 1. If $\beta_k T_{Ck} > 1$, which means that the temperature $T_k$ of the system from the point of view of our observer is less than $T_{Ck}$, we have:
\begin{equation}
\lim_{{\cal N}_k\rightarrow\infty }(2^{\beta_k T_{Ck}} - 1)^{{\cal N}_k + 2} = \infty,
\end{equation}
whereas if $\beta_k T_{Ck} < 1$, we find:
\begin{equation}
\lim_{{\cal N}_k\rightarrow\infty} (2^{\beta_k T_{Ck}} - 1)^{{\cal N}_k + 2} = 0.
\end{equation}
Hence it follows that if $T_k < T_{Ck}$, the energy is given, in effect, by an expression:
\begin{equation}
E_k(\beta_k) = \frac{2^{\beta_k T_{Ck}}}{2^{\beta_k T_{Ck}} - 2}T_{Ck}\ln(2).
\end{equation}
In particular, in the low temperature limit, where $\beta_k \longrightarrow\infty$, the energy is:
\begin{equation}
E_k = T_{Ck}\ln(2).
\end{equation}
In contrast, if $T_k > T_{Ck}$, the energy is:
\begin{equation}
E_(\beta_k) = \frac{2^{\beta_k T_{Ck}}}{2^{\beta_k T_{Ck}} - 1}({\cal N}_k + 1)T_{Ck}\ln(2),
\end{equation}
which has the property:
\begin{equation}
\lim_{T_k\rightarrow T_{Ck}^+}E_k(\beta_k) = 2({\cal N}_k + 1)T_{Ck}\ln(2).
\end{equation}

   Comparing Eqs. (7.6) and (7.9) we observe that at the characteristic temperature $T_{Ck}$ spacetime performs a {\it phase transition}, where the energy increases, from the point of view of an observer on a stretched or shrinked horizon of spacetime, by a factor ${\cal N}_k +1$. We shall see later that for macroscopic, astrophysical black holes ${\cal N}_k$ may be expected to be around $10^{80}$, whereas for the cosmological horizon of the universe ${\cal N}_k$ is around $10^{122}$. Hence there is really a tremendous increase in the energy, when the temperature exceeds $T_{Ck}$ from the point of view of our observer. Eqs. (4.15), (6.2) and (7.6) imply that when $T_k < T_{Ck}$, the horizon area is around $\ell_{Pl}^2$, where $\ell_{Pl}$ is the Planck length, and there is, in practice, no horizon at all, whereas  at the temperature $T_{Ck}$ the horizon takes a macroscopic size. In this sense the temperature $T_{Ck}$, which agrees with the Unruh temperature of the observer, may be regarded as the lowest possible temperature of spacetime from the point of view of the observer. This means that {\it all} horizons of spacetime emit thermal radiation, and the temperature of the radiation equals with the Unruh temperature of an observer with constant proper acceleration, residing close to a horizon of spacetime. The result corroborates the findings obtained in Ref. {\cite{yytoo}} by Peltola by means of a quantization of a massless scalar field close to an arbitrary horizon of a spherically symmetric spacetime.

    Defining the {\it average energy per puncture} as:
\begin{equation}
{\bar E}_k(\beta_k) := \frac{E_k(\beta_k)}{{\cal N}_k}
\end{equation}
one observes, using Eq. (7.9), that the latent heat per puncture associated with the phase transition taking place at the temperature $T_{Ck}$ is:
\begin{equation}
{\bar L}_k = 2T_{Ck}\ln(2).
\end{equation}
We also find, employing Eqs. (6.2), (6.11) and (7.9) that if we denote:
\begin{equation}
{\bar n}_k := \frac{n_k}{{\cal N}_k},
\end{equation}
then
\begin{equation}
{\bar n}_k = 2
\end{equation}
after the phase transition has been completed. This result is of particular interest, when spacetime possesses {\it just one horizon}. Spherically symmetric spacetimes possessing just one horizon include, among other things, the Schwarzschild and the de Sitter spacetimes. In such spacetimes we denote ${\bar n}:= {\bar n}_1$, and ${\bar n}$ gives the average  of the quantum numbers determining, as in Eq. (2.14), the areas contributed to the horizon by the punctures of the spin network. Since $n_p = 2j_p$, Eq. (7.13) implies that the quantum numbers $j_p$ jump, in average, from 0 to 1 in the phase transition. This means that before the phase transition all punctures on the horizon, except one, are associated with the trivial representation of the rotation group $SO(3)$, whereas after the phase transition has been completed, the punctures  are associated, in average, with the first non-trivial representation of $SO(3)$.

   For spacetimes with multiple horizons, however, the situation is more complicated. In spacetimes of that kind the area eigenvalues of the horizons are determined by the quantum numbers $n_{kp}$ and $m_{kp}$, which are related to each other as in Eq. (5.4). In that case ${\bar n}_k$ gives the average of all of the quantum numbers $n_{kp}$ and $m_{kp}$, but it is not related to the representations of the group $SO(3)$ associated with the punctures on the horizon in the same way as it is in spacetimes with just one horizon.

\section{Entropy}

In general, the entropy $S(\beta)$ of any system may be calculated from its partition function $Z(\beta)$ and average energy $E(\beta)$ by means of the formula:
\begin{equation}
S(\beta) = \beta E(\beta) + \ln Z(\beta).
\end{equation}
Using Eqs (4.15), (6.11) and (7.8) we find that when $T_k > T_{Ck}$ we have, in the leading approximation for large ${\cal N}_k$:
\begin{equation}
2^{\beta_k T_{Ck}} =  \frac{2A_k}{2A_k - A_k^{crit}},
\end{equation}
where the critical area
\begin{equation}
A_k^{crit} := 8{\cal N}_k\ln(2)
\end{equation}
is the area taken by the horizon $k$ in the limit, where $T_k\longrightarrow T_{Ck}^+$. Again, employing Eqs. (4.15) and (6.11), together with Eq. (6.12), and substituting Eq. (8.2) in Eq. (8.1) we find that in the large ${\cal N}_k$ limit we may write the entropy of spacetime as a function of the area $A_k$ of a stretched or shrinked horizon $k$ from the point of view of an observer at rest with respect to the horizon as:
\begin{equation}
S(A_k) = \frac{A_k}{4\ln(2)}\ln\left(\frac{2A_k}{2A_k - A_k^{crit}}\right) + {\cal N}_k\ln\left(\frac{2A_k - A_k^{crit}}{A_k^{crit}}\right),
\end{equation}
 or in the SI units:
\begin{equation}
S(A_k) = \frac{1}{4\ln(2)}\frac{k_B c^3}{\hbar G}A_k\ln\left(\frac{2A_k}{2A_k - A_k^{crit}}\right) + {\cal N}_k k_B\ln\left(\frac{2A_k - A_k^{crit}}{A_k^{crit}}\right).
\end{equation}

    Eq. (8.5) is valid, whenever $T_k > T_{Ck}$. A satisfactory feature of Eq. (8.5) is that the entropy depends on the horizon area $A_k$ only, being independent of the proper acceleration $a_k$ of the observer on the stretched or shrinked horizon. In the limit, where $T_k \longrightarrow T_{Ck}^+$ and hence $A_k \longrightarrow A_k^{crit+}$, we find:
\begin{equation}
S(A_k) = \frac{1}{4}\frac{k_B c^3}{\hbar G}A_k,
\end{equation}
which is the famous {\it Bekenstein-Hawking entropy law}. {\cite{kaatoo,kootoo,neetoo}} So far the physicists have been able to obtain the Bekenstein-Hawking entropy law by means of the calculation based on loop quantum gravity for the event horizons of the Schwarzschild and the Reissner-Nordstr\"om black holes only, {\cite{viitoo,kuutoo}} whereas our analysis, in turn, shows that loop quantum gravity implies the validity of the Bekenstein-Hawking entropy law from the point of view of any observer close to any horizon of any spherically symmetric spacetime. However, it should be stressed that in the classic derivations of the Bekenstein-Hawking entropy law from loop quantum gravity the horizon area spectrum is given by Eq. (1.1), rather than by Eq. (1.2), and the counting of the degenerate states is also very different from the one used in our approach. In our approach the Bekenstein-Hawking entropy law, together with its generalizations for arbitrary spherically symmetric spacetimes, comes out in a very simple and natural manner, if only the traditional area operator used in loop quantum gravity is replaced by the operator introduced by Krasnov.

  According to our analysis the Bekenstein-Hawking entropy law holds immediately after the phase transition has been completed. Nevertheless, it is easy to show that the Bekenstein-Hawking entropy law holds during the phase transition itself as well. During the phase transition $T_k = T_{Ck} = a_k/(2\pi)$ to a very good approximation, and the relationship
\begin{equation}
\beta_k = \frac{\partial S}{\partial E_k} = \frac{\partial S}{\partial A_k}\frac{dA_k}{dE_k}
\end{equation}
between the temperature parameter $\beta_k$, average energy $E_k$ and the entropy $S$ from the point of view of an observer on the stretched or shrinked horizon $k$ implies, through Eq. (4.15):
\begin{equation}
S = \frac{1}{4}\frac{k_Bc^3}{\hbar G} A_k
\end{equation}
which, again, is the Bekenstein-Hawking entropy law in Eq. (8.6).

\section{Schwarzschild Spacetime}

So far we have considered the thermodynamics of general, spherically symmetric spacetimes in loop quantum gravity. We shall now turn our attention to specific examples. The line element of vacuum spacetime involving a Schwarzschild black hole may be written as:
\begin{equation}
ds^2 = -(1 - \frac{2M}{r})\,dt^2 + \frac{dr^2}{1 - \frac{2M}{r}} + r^2\,d\theta^2 + r^2\sin^2(\theta)\,d\phi^2.
\end{equation}
Comparing Eqs. (3.1) and (9.1) we find that the only parameter is the Schwarzschild mass $M$, and we have:
\begin{equation}
f(r,M) = 1 - \frac{2M}{r}.
\end{equation}
The Schwarzschild spacetime has the event horizon, where.
\begin{equation}
r = r_S := 2M.
\end{equation}
On the event horizon 
\begin{equation}
\frac{\partial f}{\partial r}\vert_{r=r_S} = \frac{2M}{r_S^2} = \frac{1}{2M} > 0,
\end{equation}
and hence we must consider the stretched, instead of the shrinked horizon of spacetime. According to Eq. (3.7) the proper acceleration of an observer at rest with respect to the coordinates $r$, $\theta$ and $\phi$ is, whenever $r > r_S$:
\begin{equation}
a = (1- \frac{2M}{r})^{-1/2}\frac{M}{r^2},
\end{equation}
which tends to infinity, when $r\longrightarrow r_S^+$. On the stretched horizon, just outside of the event horizon, we may write the proper acceleration as:
\begin{equation}
a = B\frac{1}{4M},
\end{equation}
where
\begin{equation}
B := (1-\frac{2M}{r})^{-1/2}
\end{equation}
is the blue-shift factor.

     In our approach the thermodynamical properties of the Schwarzschild black hole stem from the quantum mechanics of the punctures of the spin network on the stretched horizon. Using Eqs. (6.11) and (9.6) we find that the characteristic temperature of the Schwarzschild black hole from the point of view of an observer at rest on a stretched horizon, where the proper acceleration $a = constant$ is:
\begin{equation}
T_C = B\frac{1}{8\pi M}.
\end{equation}
When the temperature $T$ of the Schwarzschild black hole is less than $T_C$ from the point of view of our observer, the punctures on the stretched horizon are all, except one, associated with the trivial representation of the group $SO(3)$, where $j_p = 0$, and the black hole is of the Planck size, which means that there is, in effect, no black hole at all. At the characteristic temperature $T_C$, however, the black hole performs a phase transition, where the quantum numbers $j_p$ jump, in average, from 0 to 1, and the black hole grows to a macroscopic size. In this sense the characteristic temperature $T_C$ may be regarded as the lowest possible temperature of the Schwarzschild black hole from the point of view of our observer. This means that if the temperature of the environment of the black hole is less than $T_C$, the black hole emits thermal radiation with the characteristic temperature $T_C$. If the effects of backscattering are ignored, the Tolman relation {\cite{seetoo}} implies that the temperature of the radiation far from the black hole may be obtained by means of a simple replacement $B = 1$ in Eq. (9.8). The resulting temperature is, in the SI units:
\begin{equation}
T_\infty = \frac{1}{8\pi}\frac{\hbar c^3}{Gk_B}\frac{1}{M},
\end{equation}
which agrees with the {\it Hawking temperature} {\cite{kootoo}}
\begin{equation}
T_H := \frac{1}{8\pi}\frac{\hbar c^3}{Gk_B}\frac{1}{M}
\end{equation}
of the Schwarzschild black hole. Hence we have obtained the Hawking effect from loop quantum gravity: The black hole emits thermal radiation with an effective temperature which, from the point of view of a distant observer, agrees with its Hawking temperature. We have even found a microscopic explanation to the Hawking effect: When black hole radiates, the edges at the punctures of the spin network on the stretched horizon perform, in average, a transition from the excited states, where $j_p = 1$, to the ground states, where $j_ p = 0$.
 
\section{De Sitter Spacetime}

In the static coordinates the line element of the de Sitter spacetime takes the form: {\cite{kasitoo}}
\begin{equation}
ds^2 = - (1 - \frac{\Lambda}{3}r^2)\,dt^2 + \frac{dr^2}{1 - \frac{\Lambda}{3}r^2} + r^2\,d\theta^2 + r^2\sin^2(\theta)\,d\phi^2.
\end{equation}
The de Sitter line element describes the so-called de Sitter universe, which is a spherically symmetric, homogeneous, empty universe with a positive cosmological constant $\Lambda$. One may expect that during its latter developement our Universe will become, to a very good approximation, to a de Sitter universe, whose expansion accelerates exponentially. The only parameter of the de Sitter spacetime is the cosmological constant $\Lambda$, and we have:
\begin{equation}
f(r,\Lambda) = 1 - \frac{\Lambda}{3}r^2.
\end{equation}
The de Sitter spacetime has the {\it cosmological horizon}, where
\begin{equation}
r = r_C := \sqrt{\frac{3}{\Lambda}}.
\end{equation}
On the cosmological horizon
\begin{equation}
\frac{\partial f}{\partial r}\vert_{r = r_C} = -\frac{2\Lambda}{3} r_C = -2\sqrt{\frac{\Lambda}{3}} < 0,
\end{equation}
and hence we must consider the shrinked horizon of spacetime. Eqs. (3.7) and (10.2) imply that the proper acceleration of an observer at rest with respect to the coordinates $r$, $\theta$ and $\phi$ is:
\begin{equation}
a = (1 - \frac{\Lambda}{3}r^2)^{-1/2}\frac{\Lambda}{3}r,
\end{equation}
whenever $r < r_C$. On the shrinked horizon, just inside of the cosmological horizon we must therefore have:
\begin{equation}
a = B\sqrt{\frac{\Lambda}{3}},
\end{equation}
where the blue-shift factor $B$ now takes the form:
\begin{equation}
B = (1 - \frac{\Lambda}{3}r^2)^{-1/2}.
\end{equation}

     In our approach we consider the cosmological "constant" $\Lambda$ not really as a constant, but as a free thermodynamical variable. As a consequence, $\Lambda$ may change in the thermodynamical processes. Eqs. (6.11) and (10.6) imply that the de Sitter spacetime has, from the point of view of an observer at rest on the shrinked horizon, where the proper acceleration $a = constant$, the characteristic temperature
\begin{equation}
T_C = B\frac{1}{2\pi}\sqrt{\frac{\Lambda}{3}}.
\end{equation}
This is the lowest possible temperature, which the de Sitter spacetime with macroscopic cosmological horizon may have from the point of view of our observer on the shrinked horizon. At the temperature $T_C$ the de Sitter spacetime performs a phase transition, where the quantum numbers $j_p$ associated with the edges of the spin network at its punctures on the shrinked horizon jump, in average, from 0 to 1.
 
   It is very interesting to study, what happens to the cosmological constant $\Lambda$ during the phase transition. Eqs. (4.15) and (6.11) imply that the energy of the de Sitter spacetime from the point of view of an observer on the shrinked horizon may be written as:
\begin{equation}
E = \frac{1}{4}T_CA_C,
\end{equation}
 where
\begin{equation}
A_C := 4\pi r_C^2 = \frac{12\pi}{\Lambda}
\end{equation}
is the area of the cosmological horizon. Employing Eqs. (7.2) and (7.3), together with Eqs. (10.9) and (10.10) we find that the cosmological constant $\Lambda$ takes, when written as a function of the temperature parameter $\beta$, the form:
\begin{equation}
\Lambda(\beta) = \frac{3\pi}{\ln(2)}\left[\frac{2^{\beta T_C}}{2^{\beta T_C} - 2} + \frac{2^{\beta T_C}}{2^{\beta T_C} - 1 - (2^{\beta T_C} - 1)^{N+2}}(N+1)\right]^{-1},
\end{equation}
where $N$ is the number of the punctures of the spin network on the shrinked horizon. Assuming that $N$ is very large, we observe that the cosmological constant has the property:
\begin{equation}
\lim_{\beta\rightarrow\infty}\Lambda(\beta) = \frac{3\pi}{\ln(2)}
\end{equation}
or, when written in the SI units:
\begin{equation}
\lim_{\beta\rightarrow\infty}\Lambda(\beta) = \frac{3\pi}{\ln(2)}\frac{c^2}{\ell_{Pl}^2} \approx 4.8\times 10^{87} s^{-2}.
\end{equation}
The number on the right hand side of Eq. (10.13), which gives the cosmological constant in the low-temperature limit, is really huge. Actually, the Planck-scale value given by the right hand side of Eq. (10.13) for the cosmological constant is exactly the value one expects on the dimensional grounds. Whenever the temperature $T < T_C$, the cosmological constant is given, in effect, by an expression:
\begin{equation}
\Lambda(\beta) = \frac{3\pi}{\ln(2)}\frac{2^{\beta T_C} - 2}{2^{\beta T_C}}.
\end{equation}
At the characteristic temperature $T_C$, however, the cosmological constant drops drastically. When $T > T_C$, the effective dependence of the cosmological constant on the temperature parameter $\beta$ becomes to:
\begin{equation}
\Lambda(\beta) = \frac{3\pi}{\ln(2)}\frac{1}{N+1}\frac{2^{\beta T_C} - 1}{2^{\beta T_C}},
\end{equation}
and we find that, when written in the SI units: 
\begin{equation}
\lim_{T\rightarrow T_C^+}\Lambda(\beta) = \frac{3\pi}{2\ln(2)}\frac{1}{N+1}\frac{c^2}{\ell_{Pl}^2}.
\end{equation}

    The right and side of Eq. (10.16) gives the cosmological constant after the phase transition has been completed. Putting
\begin{equation}
N \sim 10^{122}
\end{equation}
in Eq. (10.16) we get:
\begin{equation}
\lim_{T\rightarrow T_C^+} \Lambda(\beta) \sim 10^{-35}s^{-2},
\end{equation}
which is the current estimate, based on the observations, for the cosmological constant. Hence loop quantum gravity provides a potential explanation both to the presence and the smallness of the cosmological constant: The integer $N$ in Eq. (10.16) is finite, even though admittedly very large, and therefore the cosmological constant is non-zero. On the other hand, however, the largeness of $N$ implies that the value taken by the cosmological constant after the phase transition is very small. The reason for the smallness of the cosmological constant is, quite simply, that the universe is a far too hot place to maintain large cosmological constant. Putting the blue-shift factor $B = 1$ in Eq. (10.8) we find that the cosmological horizon emits, from the point of view of an observer far from the horizon, thermal radiation with a characteristic temperature
\begin{equation}
T_\Lambda = \frac{1}{2\pi}\sqrt{\frac{\Lambda}{3}}
\end{equation}
which, when written in the SI units, takes the form:
\begin{equation}
T_\Lambda = \frac{1}{2\pi}\sqrt{\frac{\Lambda}{3}}\frac{\hbar}{k_B}\sim 10^{-30}K.
 \end{equation}
Eq. (10.20) reproduces the result obtained by Gibbons and Hawking in Ref. {\cite{ysitoo}}. As one may observe, the temperature $T_\Lambda$ is vey low, indeed. It is possible that at the moment of the creation of the Universe the cosmological constant took a huge, Planck-size value of Eq. (10.13). However, almost immediately after its creation the universe performed a phase transition, where the cosmological constant dropped to its present, very small value. In this sense the smallness of the cosmological constant may be regarded as a successful prediction of loop quantum gravity. {\cite{kakskyt}}

   The key role in our discussion about the properties of the cosmological constant was played by the phase transition, which takes place, from the point of view of the observer on the shrinked horizon of the de Sitter spacetime, at the characteristic temperature $T_C$. The presence of such phase transition followed from  our decision to replace the standard area operator used in loop quantum gravity by the area operator introduced by Krasnov. According to the best knowledge of the author, no such phase transition has been reported in those approaches to loop quantum gravity, which have been based on the use of the standard area operator. Hence it appears that the phase transition found in this paper is an entirely novel feature of loop quantum gravity, and the interesting as well as physically sensible consequences of its presence may be viewed as an argument for an idea that perhaps it is the operator introduced by Krasnov, rather than the standard area operator, which should be used as the area operator in loop quantum gravity. 

      An interesting approach to the problem of cosmological constant in loop quantum gravity, based on the use of spin foams, has been put forward by Dupuis and Girelli in Ref. {\cite{kakskyta}}. In this approach the group $SU(2)$ in loop quantum gravity was replaced by a quantum deformed group ${\cal U}_q(su(2))$. Using this quantum group Dupuis and Girelli were able to construct the length- and the area operators in loop quantum gravity, and they found that in 3 spacetime dimensions the resulting geometry is hyperbolic, thus resembling the geometry of the de Sitter spacetime. (For other spin foam-related approaches, see Refs. {\cite{kakskytb, kakskytc}}).

\section{Reissner-Nordstr\"om Spacetime}

The Reissner-Nordstr\"om spacetime is a spherically symmetric, asymptotically flat spacetime involving the Reissner Nordstr\"om black hole. Its geometry is described by the line element:
\begin{equation}
ds^2 = -(1 - \frac{2M}{r} + \frac{Q^2}{r^2})\,dt^2 + \frac{dr^2}{1 - \frac{2M}{r} + \frac{Q^2}{r^2}} + r^2\,d\theta^2 + r^2\sin^2(\theta)\,d\phi^2.
\end{equation}
The Reissner-Nordstr\"om spacetime has two parameters, which are the masss $M$ and the electric charge $Q$ of the Reissner-Nordstr\"om black hole. We have:
\begin{equation}
f(r,M,Q) = 1 - \frac{2M}{r} + \frac{Q^2}{r^2}.
\end{equation}
Even though the Reissner-Nordstr\"om black hole solution is probably unphysical, {\cite{kaayy}} it nevertheless provides an interesting example of a solution to the combined Einstein-Maxwell equations, where spacetime has {\it two}, and not just one horizon. The horizons of the Reissner-Nordstr\"om spacetime are located at those two-spheres, where 
\begin{equation}
r = r_{\pm} := M \pm \sqrt{M^2 - Q^2}.
\end{equation}
At the {\it inner horizon} $r = r_{-}$, and at the {\it outer horizon}, which is the true event horizon of the hole, $r = r_{+}$. For an extreme black hole these two horizons coincide, and we have $M = \vert Q\vert$. 

    Using Eqs. (11.2) and (11.3) one finds that
\begin{subequations}
\begin{eqnarray}
\frac{\partial f}{\partial r}\vert_{r=r_{-}} &=&\frac{2}{r^3_{-}}(Mr_-  - Q^2)= -\frac{2}{r_-^2}\sqrt{M^2 - Q^2} < 0,\\
\frac{\partial f}{\partial r}\vert_{r = r_+} &=&\frac{2}{r_+^3}( Mr_+ - Q^2) = \frac{2}{r_+^2}\sqrt{M^2 - Q^2} > 0.
\end{eqnarray}
\end{subequations}
Hence we must consider the shrinked horizon of the inner horizon, and the stretched horizon of the outer horizon. Eq. (3.7) implies that the proper acceleration of an observer at rest with respect to the coordinates $r$, $\theta$ and $\phi$ is:
\begin{equation}
a = (1 - \frac{2M}{r} + \frac{Q^2}{r^2})^{-1/2}\vert \frac{M}{r^2} - \frac{Q^2}{r^3}\vert,
\end{equation}
whenever we have either $r < r_-$ or $r > r_+$. Using Eqs. (6.11) and (11.3) we find that the characteristic temperature perceived by an observer on the stretched horizon $a= constant$ just outside of the outer horizon is
\begin{equation}
T_+ = B\frac{\kappa_+}{2\pi},
\end{equation}
whereas that perceived by an observer on the shrinked horizon $ a = constant$ just inside of the inner horizon is
\begin{equation}
T_- = B\frac{\kappa_-}{2\pi}.
\end{equation}
In Eqs. (11.6) and (11.7)
\begin{equation}
B := (1 - \frac{2M}{r} + \frac{Q^2}{r^2})^{-1/2}
\end{equation}
is the blue-shift factor,
\begin{equation}
\kappa_+ := \frac{\sqrt{M^2 - Q^2}}{r_+^2}
\end{equation}
is the surface gravity at the outer horizon, and
\begin{equation}
\kappa_- := \frac{\sqrt{M^2 - Q^2}}{r_-^2}
\end{equation}
the (modulus of the) surface gravity at the inner horizon.

       The temperatures $T_+$ and $T_-$, respectively, are the lowest possible temperatures, which the Reissner-Nordstr\"om black hole may have from the points of views of observers on the stretched and the shrinked horizons. At the temperature $T_-$ the Reissner-Nordstr\"om spacetime performs, from the point of view of an observer on the shrinked horizon, a phase transition, where the quantum numbers $j_p$ associated with the punctures of the spin network on the shrinked horizon jump, in average, from 0 to 1. Unfortunately, no such simple microscopic interpretation may be given for the phase transition, which takes place, from the point of view of an observer on the stretched horizon, when the temperature $T = T_+$. However, in the special case, where the numbers $N_1$ and $N_2$ of the punctures of the spin network on the shrinked and the stretched horizons, respecteively, are equals, {\it i. e.} $N_1 = N_2$, a simple interpretation exists: If $N_1 = N_2$, then ${\cal N}_2 = 2N_2$, and if we denote by $n_2$ the quantum number determining the area $A_2$ of the stretched horizon such that
\begin{equation}
A_2  = 4\ln(2)\ell_{Pl}^2 n_2,
\end{equation}
then Eqs. (7.12) and (7.13) imply:
\begin{equation}
\frac{n_2}{N_2} = 4
\end{equation}
immediately after the phase transition has been completed. This means that the quantum numbers $j_p$ associated with the punctures on the stretched horizon jump, in average, from 0 to 2 during the phase transition. In general, if $N_2 > N_1$, then Eq. (7.13) implies:
\begin{equation}
\frac{n_2}{N_2} = \frac{n_2}{{\cal N}_2}\frac{{\cal N}_2}{N_2} = 2\frac{{\cal N}_2}{N_2} = 2\frac{N_1 + N_2}{N_2} = 2(1 + \frac{N_1}{N_2})
\end{equation}
after the completion of the phase transition. Hence it follows that the average jump in the quantum numbers $j_p$ on the stretched horizon  is
\begin{equation}
\Delta{\bar j}_p = 1 + \frac{N_1}{N_2}.
\end{equation}

    The fact that we may associate a certain minimum temperature not only with the outer horizon, but also with the inner horizon of the Reissner-Nordstr\"om spacetime indicates that {\it both} of the horizons of the Reissner-Nordstr\"om black hole will radiate: The outer horizon emits radiation outwards, whereas the inner horizon emits radiation inwards. This result confirms the findings made in Ref. {\cite{kaakaa}}, where a detailed analysis, based on the use of the quantum field theory in curved spacetime, on the properties of the radiation emitted by the inner horizon was performed. One should not be surprised by the difference in the temperatures measured for the radiation on the stretched and the shrinked horizons, because the inner, and the outer horizons of the Reissner-Nordstr\"om black hole are not in a causal contact with each other. Note, however, that according to Eqs. (11.9) and (11.10) both of those temperatures are zero, when the hole is extreme, {\it i. e.} $M = \vert Q\vert$.

   It is interesting to study the spectra of the mass $M$ and the electric charge $Q$ of the Reissner-Nordstr\"om black hole in our model. The area of the outer horizon of the Reissner-Nordstr\"om black hole is:
\begin{equation}
A_+ = 4\pi r_+^2,
\end{equation}
whereas the area of its inner horizon is
\begin{equation}
A_- = 4\pi r_-^2.
\end{equation}
For all practical purposes we may identify the areas of the stretched and the shrinked horizons of the Reissner-Nordstr\"om black hole, respectively, with the areas of the outer, and the inner horizons of the hole. Hence we have:
\begin{subequations}
\begin{eqnarray}
A_ - &=& 4\ln(2)\ell_{Pl}^2n_1,\\
A_+ &=& 4\ln(2)\ell_{Pl}^2n_2,
\end{eqnarray}
\end{subequations}
where $n_1$ and $n_2$ are positive integers such that $n_2 \ge n_1$. Employing Eqs. (11.3), (11.15) and (11.16) we find that the possible values of the mass $M$ are of the form:
\begin{equation}
M = \frac{1}{2}\sqrt{\frac{\ln(2)}{\pi}}(\sqrt{n_1} + \sqrt{n_2}),
\end{equation}
whereas those of the electric charge $Q$ are of the form:
\begin{equation}
Q = \pm\sqrt{\frac{\ln(2)}{\pi}}(n_1n_2)^{1/4}.
\end{equation}
The spectra of $M$ and $Q$ are somewhat different from those in, for instance, Refs. {\cite{kaakoo}} and {\cite{seite}}, where $Q^2$ had a uniformly spaced spectrum. For an extreme black hole, however, $n_ 1 = n_2$, and the spectrum of $Q^2$, as well as that of $M^2$, are uniformly spaced.

 \section{Kottler Spacetime}

In the Schwarzschild-de Sitter spacetime, which is sometimes also known as the {\it Kottler spacetime}, the line element reads as:
\begin{equation}
ds^2 = - (1 - \frac{2M}{r} - \frac{\Lambda}{3}r^2)\,dt^2 + \frac{dr^2}{1 - \frac{2M}{r} - \frac{\Lambda}{3}r^2} + r^2\,d\theta^2 + r^2\sin^2(\theta)\,d\phi^2.
\end{equation}
The Kottler spacetime has two parameters, which are the mass $M$ and the  positive cosmological constant $\Lambda$. We have:
\begin{equation}
f(r,M,\Lambda) = 1 - \frac{2M}{r} - \frac{\Lambda}{3}r^2.
\end{equation}
As one may observe, the Kottler spacetime reduces to the Schwarzschild spcaetime, when $\Lambda = 0$, and to the de Sitter spacetime, when $M = 0$. Eq. (12.2) implies that the horizons of the Kottler spacetime are those spacelike two-spheres, where
\begin{equation}
-\frac{\Lambda}{3}r^3 + r - 2M = 0.
\end{equation}
Using the standard formulas {\cite{kaanee}} for the solutions of the cubic equation one finds that Eq. (12.3) has at most two positive, real solutions. These solutions are:
\begin{subequations}
\begin{eqnarray}
r  &=& r_+ := \frac{2\sqrt{3}}{3}r_C\cos(\frac{\varphi}{3}),\\
r &=& r_- := \frac{2\sqrt{3}}{3}r_C\cos(\frac{\varphi}{3} + \frac{4\pi}{3}),
\end{eqnarray}
\end{subequations}
where
\begin{equation}
\cos(\varphi) = -\frac{3\sqrt{3}}{2}\frac{r_S}{r_C}.
\end{equation}
In Eqs. (12.4) and (12.5)
\begin{equation}
r_C := \sqrt{\frac{3}{\Lambda}}
\end{equation}
is the radius of the cosmological horizon of the resulting de Sitter spacetime, when $M=0$, and 
\begin{equation}
r_S := 2M
\end{equation}
is the radius of the Schwarzschild horizon of the resulting Schwarzschild spacetime, when $\Lambda = 0$. Eq. (12.5) implies that we must necessarily have
\begin{equation}
\frac{r_S}{r_C} \le \frac{2\sqrt{3}}{9}.
\end{equation}
If the condition (12.8) does not hold, there will be no horizon at all. In the special case, where
\begin{equation}
\frac{r_S}{r_C} = \frac{2\sqrt{3}}{9},
\end{equation}
the two horizons will coincide, and we have:
\begin{equation}
r_+ = r_- = \frac{2\sqrt{3}}{3}r_C\cos(\frac{\pi}{3}) = \frac{\sqrt{3}}{3}r_C.
\end{equation}

   Eq. (12.5) implies that $\cos(\varphi) < 0$, and hence we may pick up the angle $\varphi$ such that
\begin{equation}
\frac{\pi}{2} < \varphi \le \pi.
\end{equation}
So we find that
\begin{equation}
\frac{1}{2} = \cos(\frac{\pi}{3}) \le \cos(\frac{\varphi}{3}) < \cos(\frac{\pi}{6}) = \frac{\sqrt{3}}{2}.
\end{equation}
We also find:
\begin{equation}
0 = \cos(\frac{3\pi}{2}) < \cos(\frac{\varphi}{3} + \frac{4\pi}{3}) \le \cos(\frac{\pi}{3}) = \frac{1}{2}, 
\end{equation}
and therefore
\begin{equation}
r_+ \ge r_-.
\end{equation}
Because of that we shall call the horizon, where $r = r_+$ as the {\it outer horizon}, and the horizon, where $r = r_-$ as the {\it inner horizon} of the Kottler spacetime.
Using Eqs. (12.2), (12.4) and (12.5) we find that
\begin{subequations}
\begin{eqnarray}
\frac{\partial f}{\partial r}\vert_{r=r_+} &=& \frac{\sqrt{3}}{2 r_C}\frac{1 - 4\cos^2(\frac{\varphi}{3})}{\cos(\frac{\varphi}{3})},\\
\frac{\partial f}{\partial r}\vert_{r = r_-} &=& \frac{\sqrt{3}}{2r_C}\frac{1 - 4\cos^2(\frac{\varphi}{3} + \frac{4\varphi}{3})}{\cos(\frac{\varphi}{3} + \frac{4\pi}{3})}.
\end{eqnarray}
\end{subequations}
Employing Eqs. (12.12) and (12.13) we observe that:
\begin{subequations}
\begin{eqnarray}
\frac{\partial f}{\partial r}\vert_{r=r_+} &\le& 0,\\
\frac{\partial f}{\partial r}\vert_{r=r_-} &\ge& 0.
\end{eqnarray}
\end{subequations}
Hence we must consider a stretched horizon just outside of the inner horizon, where $r = R_-$, and a shrinked horizon just inside of the outer horizon, where $r = r_+$. In general, the proper acceleration of an observer at rest with respect to the coordinates $r$, $\theta$ and $\phi$ is, according to Eq. (3.7):
\begin{equation}
a = (1  - \frac{2M}{r} - \frac{\Lambda}{3}r^2)^{-1/2}\vert \frac{M}{r^2} - \frac{\Lambda}{3}r\vert,
\end{equation}
whenever $r_- < r < r_+$. So we find that an observer on a shrinked horizon $a = constant$ just inside of the outer horizon observes thermal radiation with a characteristic temperature
\begin{equation}
T_+ = B\frac{\sqrt{3}}{8\pi r_C}\frac{4\cos^2(\frac{\varphi}{3}) - 1}{\cos(\frac{\varphi}{3})},
\end{equation}
whereas an observer on a stretched horizon $a = constant$ just outside of the inner horizon observes thermal radiation with temperature
\begin{equation}
T_- = B\frac{\sqrt{3}}{8\pi r_C}\frac{1 - 4\cos^2(\frac{\varphi}{3} + \frac{4\pi}{3})}{\cos(\frac{\varphi}{3} + \frac{4\pi}{3})}.
\end{equation}
In Eqs. (12.18) and (12.19)
\begin{equation}
B := (1 - \frac{2M}{r} - \frac{\Lambda}{3}r^2)^{-1/2}
\end{equation}
is the blue-shift factor.

     It is interesting to study, which form the temperatures $T_+$ and $T_-$ will take in the limit, where
\begin{equation}
\frac{r_S}{r_C} \ll 1.
\end{equation}
In this limit Eq. (12.5) implies that we may write:
\begin{equation}
\varphi = \frac{\pi}{2} + \epsilon,
\end{equation}
where the positive angle $\epsilon \ll 1$. For instance, if we have a supermassive black hole, whose mass is $10^9$ solar masses, or $10^{39}kg$, then $r_S \sim 10^{12}m$, and if we put $\Lambda \sim 10^{-35}s^{-2}$, we have $r_C \sim 10^{26}m$, which means that $r_S/r_C \sim 10^{-14}$. In terms of the angle $\epsilon$ we may write the temperatures $T_+$ and $T_-$ as:
\begin{subequations}
\begin{eqnarray}
T_+ &=& B\frac{1}{2\pi}\frac{1}{r_C} + O(\epsilon),\\
T_- &=& B\frac{3\sqrt{3}}{8\pi}\frac{1}{r_C}\frac{1}{\epsilon} + O(\epsilon),
\end{eqnarray}
\end{subequations}
where $O(\epsilon)$ denotes the terms proportional to the first or higher powers of $\epsilon$. On the other hand, Eq. (12.5) implies:
\begin{equation}
\frac{3\sqrt{3}}{2}\frac{r_S}{r_C} = \epsilon + O(\epsilon^2),
\end{equation}
where $O(\epsilon^2)$  denotes the terms proprtional to the second or higher powers of $\epsilon$. Hence we find that when Eq. (12.21) holds, we may write, in effect:
\begin{subequations}
\begin{eqnarray}
T_+ &=& B\frac{1}{2\pi}\frac{1}{r_C} = B\frac{1}{2\pi}\sqrt{\frac{\Lambda}{3}},\\
T_- &=& B\frac{1}{4\pi}\frac{1}{r_S} = B\frac{1}{8\pi M},
\end{eqnarray}
\end{subequations}
Far from the both horizons we have $B=1$, and if the backscattering effects of the radiation from the spacetime geometry are ignored, we have, in the SI units:
\begin{subequations}
\begin{eqnarray}
T_+ &=& \frac{1}{2\pi}\sqrt{\frac{\Lambda}{3}}\frac{\hbar}{k_B},\\
T_- &=& = \frac{1}{8\pi}\frac{\hbar c^3}{Gk_B}\frac{1}{M}.
\end{eqnarray}
\end{subequations}
As one may onserve, the temperature $T_+$ agrees with the temperature $T_\Lambda$ in Eq. (10.20), whereas the temperature $T_-$ agrees with the Hawking temperature $T_H$ in Eq. (9.10). So we may conclude that when Eq. (12.21) holds, the temperature of the radiation emitted by the outer horizon inside of the horizon agrees with the temperature of the radiation emitted by the cosmological horizon of the de Sitter spacetime, whereas the temperature of the radiation emitted outwards by the inner horizon agrees with the temperatre of the radiation emitted by the Schwarzschild black hole. The result is exactly what one expects. 

            Reasoning similar to the one performed in Section 11 for the Reissner-Nordstr\"om spacetime indicates that when the inner horizon emits radiation in the limit, where Eq. (12.21) holds,  the Kottler spacetime performs a phase transition, where the quantum numbers $j_p$ associated with the punctures of the spin network on the stretched horizon jump, in average, from 1 to 0. Likewise, when the outer horizon emits radiation inwards, the quantum numbers $j_p$ associated with the punctures of the spin network on the shrinked horizon jump, in average, from
\begin{equation}
{\bar j}_p := 1 + \frac{N_1}{N_2}
\end{equation}
to 0. In Eq. (12.24) $N_1$ and $N_2$, respectively, are the numbers of the punctures on the stretched and on the shrinked horizons. Assuming that Eq. (12.21) holds, we must have $N_2 \gg N_1$, and hence we have, in effect, ${\bar j}_p = 1$. In other words, during the emission of the radiation $j_p$ jumps, in average, from 1 to 0 in the punctures located both on the stretched and on the shrinked horizons of the Kottler spacetime.

  One of the essential differences between the Kottler and the Reissner-Nordstr\"om spacetimes is that in the Kottler spacetime the two horizons are in a causal contact with each other, whereas in the Reissner-Nordstr\"om spacetime they are not. As a consequence, a horizon of the Kottler spacetime does not really begin to emit radiation unless its temperature, from the point of view of an observer, is less than the temperature of the remaining horizon. For instance, if the temperature of the outer horizon is less than the temperature of the inner horizon, the inner horizon emits, and the outer horizon absorbs radiation. When  the inner horizon radiates , the quantum numbers $j_p$ associated with the punctures of the spin network on the corresponding stretched horizon will decrease, and the inner horizon shrinks. At the same time the quantum numbers $j_q$ at the punctures $q$ of the spin network on the shrinked horizon just inside of the outer horizon will increase, and the outer horizon expands. The thermal equilibrium between the two horizons is never reached, because the smaller becomes the inner horizon, the higher becomes its temperature, whereas the temperature of the outer horizon becomes lower, when it size is increased.

\section{Concluding Remarks}

   In this paper we have investigated the thermodynamics of general, spherically symmetric spacetimes in loop quantum gravity using the operator introduced by Krasnov in Ref. {\cite{nee}} as the area operator. The most important feature of Krasnov's area operator is that in contrast to the standard area operator used in loop quantum gravity, the spectrum of this area operator is uniformly spaced. A disadvantage of the area operator is that it is not $SO(3)$ invariant. However, at the horizons of spacetime the local $SO(3)$ invariance of loop quantum gravity does not necessarily hold, and the operator introduced by Krasnov provides a legitimate area operator. The simplicity of the resulting horizon area spectrum allows one to obtain an explicit, analytic expression for the partition function of the spacetime from the point of view of an observer with a constant proper acceleration close to a horizon of spacetime. Among other things, the partition function implied that spacetime has a certain minimum temperature, which agrees with the Unruh temperature of the observer. At this temperature the entropy of spacetime is, in the natural units, one-quarter of the horizon area. After considering the thermodynamics of the spherically symmetric spacetimes in the general level we applied our results to the Schwarzschild and the de Sitter spacetimes, which both have just one horizon, and to the Reissner-Nordstr\"om and the Kottler spacetimes, which have two horizons.

    Our decision to consider the thermodynamics of spacetime from the point of view of an observer with a constant proper acceleration was based on an idea that observers with the same proper acceleration may be viewed in some sense equivalent. We assumed that an observer lies on a spacelike two-sphere with radius $r$ either just outside, or just inside of the horizon such that when the parameters of spacetime (mass, electric charge, cosmological constant, etc.) are changed, the radius $r$ of the two-surface will also change, but in such a way that the proper acceleration $a$ of the observer preserves unchanged. A two-surface of this kind lying just outside of the horizon itself was called as a {\it stretched horizon}, whereas a two-sphere lying just inside of the horizon was called as a {\it shrinked horizon}.

   One of the most important conclusions of our analysis was that {\it all} horizons of spacetime emit radiation. The inner horizon of the Reissner-Nordstr\"om spacetime, for instance, emits radiation inwards, and its outer horizon outwards, whereas in the Kottler spacetime the inner horizon emits outwards, and the outer horizon inwards. In loop quantum gravity the radiation of horizons may be understood, according to our approach,  as an outcome of a {\it phase transition} taking place at the punctures of the spin network on the stretched and the shrinked hroizons of spacetime. When spacetime has just one horizon, the mechanism of this phase transition, in the microscopic level, is very simple. At the onset of the Hawking radiation emitted by the Schwarzschild black hole, for instance, each puncture $p$ of the spin network on the stretched horizon of the hole is associated with a certain $(2j_p + 1)$-dimensional  irreducible representation of the rotation group $SO(3)$. When the radiation begins, we have, in average, $j_p = 1$ at each puncture. In other words, the punctures  of the stretched horizon are, in average, associated with the first non-trivial representation of the group $SO(3)$. During the radiation the non-trivial representations are being replaced by the trivial representations, where $j_p = 0$.  Finally, the representations at all punctures, except at one, have been replaced by the trivial representations, and we are left with a Planck-size remnant of the hole. All the time, when the hole radiates, the temperature of the radiation remains the same from the point of view of our observer. For spacetimes with multiple horizons the overall process is somewhat similar, but its details are a bit more complicated.

    As we have seen, loop quantum gravity provides, when the standard area operator is replaced by the one introduced by Krasnov, a microscopic explanation for the entropy and the radiation of the black holes. Moreover, the entropic and the radiative properties of the black holes may be obtained from loop quantum gravity by means of a staightforward and surprisingly simple reasoning. The most striking predictions of loop quantum gravity, however, concern the properties of the cosmological constant. If one applies the general results obtained for spherically symmetric spacetimes to the de Sitter spacetime one finds, keeping the cosmological "constant" $\Lambda$ not really as a constant, but as a free thermodynamical variable, that loop quantum gravity provides an explanation both to the presence and the smallness of the cosmological constant. For very low temperatures our approach predicts for the cosmological constant $\Lambda$ a huge, Planck-scale value which, in the SI units, is around $10^{87}s^{-2}$. However, when the temperature of the de Sitter spacetime reaches a temperature which, from the point of view of an observer on the shrinked horizon, equals with the Unruh temperature of the observer, spacetime performs a phase transition. In this phase transition the quantum numbers $j_p$ determining the representations of the group $SO(3)$ associated with the punctures of the spin network on the shrinked horizon jump, in average, from 0  to 1. As a consequence, the radius of the cosmological horizon of the de Sitter spacetime increases enormously, which means that the cosmological constant $\Lambda$ drops drastically. The resulting value of $\Lambda$ is inversely proprotional to the number $N$ of the punctures of the spin network on the shrinked horizon, and since $N$ is finite, the cosmological constant is bounded from below. However, $N$ is also very large, and therefore the cosmological constant is very small. Choosing $N \sim 10^{122}$ one finds that $\Lambda \sim 10^{-35}s^{-2}$, which is the present estimate, based on the observations, for the cosmological constant. Hence the explanation provided by loop quantum gravity to the observed smallness of the cosmological constant is, quite simply, that the present universe is a far hot place to maintain large cosmological constant. It is possible that at the moment of the creation of the universe the cosmological constant took a huge, Planck-size value. However, almost immediately after its creation  the universe performed a phase transition, where the cosmological constant dropped enormously, and the matter and the radiation of the universe were created out of the huge vacuum energy associated with the large cosmological constant. {\cite{kaavii}}

    It is really amazing, how a minor modification of the area operator generally used in loop quantum  gravity brings along far-reaching consequences in cosmology and in the physics of black holes. Given that loop quantum gravity really follows from a straightforward application of the standard rules of quantum mechanics to Einsteinä's general theory of relativity, making no other assumptions, one may expect its predictions to be pretty reliable. Even though we have restricted our attention, in this paper, to spherically symmetric spacetimes,  results similar to those obtained in this paper will most likely hold in other types of spacetime as well. For instance, it should be possible to derive the entropic and the radiative properties of the Kerr-Newman black holes by means of an analysis, which is almost identical to the one performed in this paper for the Reissner-Nordstr\"om black holes. The general results obtained in this paper for spherically symmetric spacetimes may also be applied to various cosmological models, provided that spacetime in those models possesses horizons.


\begin{thebibliography}              {30}
\bibitem{kakskyt} The idea that the properties of the cosmological constant may arise from the microscopic structure of spacetime has been advocated, for example, by T. Padmanabhan and H. Padmanabhan, Int. J. Mod. Phys. {\bf D23}, 1430011 (2014).

\medskip
\end{thebibliography}
\end{document}